\documentclass[a4paper,11pt]{article}
\usepackage[left=2.5cm,right=2.5cm,top=2cm,bottom=2.5cm]{geometry}
\usepackage{epsfig,amsfonts,amssymb}
\usepackage{xcolor}
\usepackage{amsmath}
\usepackage{cite}
\usepackage{bm}
\usepackage[normalem]{ulem}
\usepackage{framed}
\usepackage{varioref}
\colorlet{shadecolor}{lightgray}
\usepackage{hyperref}
\usepackage{mathrsfs}
\usepackage{parskip}

\usepackage[utf8]{inputenc}
\usepackage{soul}

\definecolor{blue}{rgb}{0.0, 0.0, 1.0}

\def \be {\begin{equation}}
\def \ee {\end{equation}}
\def \bea {\begin{eqnarray}}
\def \eea {\end{eqnarray}}

\labelformat{equation}{Eq.~(#1)} 

\begin{document}

\baselineskip 24pt

\begin{center}

{\Large \bf Soft factors with AdS radius corrections}

\end{center}

\vskip .6cm
\medskip

\vspace*{4.0ex}

\baselineskip=18pt

\centerline{\large \rm Karan Fernandes$^{a,b}$, Nabamita Banerjee$^{c}$ and Arpita Mitra$^{d}$}

\vspace*{4.0ex}

\centerline{\large \it ~$^a$Department of Physics, \\
National Taiwan Normal University, Taipei, 11677, Taiwan}

\centerline{\large \it ~$^b$Center of Astronomy and Gravitation, National Taiwan Normal University, Taipei 11677, Taiwan}
\centerline{\large \it ~$^c$ Indian Institute of Science Education \& Research Bhopal,}
\centerline{\large \it ~Bhopal Bypass Road, Bhauri, Bhopal 420 066, Madhya Pradesh, India}

\centerline{\large \it ~$^d$ Department  of  Physics,  Pohang  University  of  Science  and  Technology,  Pohang  37673,  Korea}

\vspace*{1.0ex}
\centerline{\small E-mail: karanfernandes86@gmail.com, nabamita@iiserb.ac.in, arpitamitra89@gmail.com}

\vspace*{5.0ex}

\centerline{\bf Abstract} \bigskip

 We review recent developments concerning the soft factorization of scattering amplitudes that arise in the large radius limit of four dimensional Anti-de Sitter (AdS$_4$) spacetimes. This includes the presence of AdS radius dependent corrections of known flat spacetime soft factors and their implication on the relationship between soft theorems and Ward identities of the boundary conformal field theory.  
 \\~~\\~~\\
 
~~
\vspace{20em}

Contribution to the Proceedings of the $6^{\text{th}}$ International Conference on Holography, String Theory and Spacetime, Da Nang, Vietnam, February 2023

\vfill \eject

\baselineskip 18pt

\tableofcontents
\section{Introduction}

Infrared properties of scattering amplitudes on asymptotically flat spacetimes have garnered considerable interest over the past decade in large part due to their relationship with asymptotic symmetries~\cite{Strominger:2017zoo,Strominger:2013jfa,Kapec:2014zla,He:2014cra,Cachazo:2014fwa,Campiglia:2015yka, McLoughlin:2022ljp}. Of central importance in these developments are soft theorems, which describe the factorization of scattering amplitudes when one or more of the particles have a vanishing energy, or soft, limit. For concreteness, we consider the case of an $n+1$ particle amplitude ${\cal{M}}_{n+1}$ amplitude comprising of $n$ hard particles with momenta $p_n$ and a single soft particle with momentum $k$ and polarization $\varepsilon$ in $D \ge 4$ spacetime dimensions. The soft theorem then informs us that in the $k \to 0$ limit we have

\begin{equation}
{\cal{M}}_{n+1}(\varepsilon\,,k\,; p_n) = S \, {\cal{M}}_{n}(p_n)
\label{st.gen}
\end{equation}

where $S$ is the soft factor that depend on all the momenta in the process and the polarization of the soft particle, while ${\cal{M}}_{n}$ is the hard particle amplitude excluding the soft particle. For the remainder of this paper, we work in $D=4$ spacetime dimensions, where properties of soft theorems and their relationship with asymptotic symmetries are best understood. On expressing the soft particle momentum as $k = \omega(1\,, \hat{k})$, where $\omega$ is the frequency and $\hat{k}$ the orientation, we generally have the following expansion of $S$ in the $\omega \to 0$ limit

\begin{equation}
S = \frac{1}{\omega} A_0 + \ln \omega^{-1} A_1 + A_2 + \mathcal{O}(\omega),,
\label{sf.gen}
\end{equation}

with the $A_i$'s ($i = 0,1,\cdots$) being specific to the soft particle considered and the hard particle amplitude. In this regard, Weinberg~\cite{Weinberg:1964ew,Weinberg:1965nx} showed that the leading divergent term in \ref{sf.gen} takes on a universal form for the soft photon and graviton cases for all gauge and diffeomorphism invariant theories, regardless of the amplitude for the hard particle. This follows from the gauge invariance of the leading soft photon factor being a consequence of charge conservation for the hard particles, and likewise diffeomorphism invariance of leading soft graviton factor is a consequence of a conserved stress energy tensor. The indicated subleading terms in \ref{sf.gen} are also universal in the soft graviton case for all amplitudes, and the diffeomorphism invariance of these terms is a consequence of angular momentum conservation \cite{Cachazo:2014fwa, McLoughlin:2022ljp}. The higher order subleading contributions contained in $\mathcal{O}(\omega)$ are generally theory specific and involve loop corrections from the amplitude.

The importance in better understanding the soft expansion comes from two broad class of developments on asymptotically flat spacetimes. The first of these concern a web of relations known as the `infrared triangle' in $D=4$ spacetime dimensions~\cite{Strominger:2017zoo}. The leading soft theorem (with $S$ in \ref{st.gen} involving the leading divergent term of \ref{sf.gen}) can remarkably be derived from a symmetry principle. Gauge and gravitational fields can involve large gauge transformations that survive across null infinity, and soft theorems for the corresponding fields are equivalent to large gauge Ward identities satisfied by the $S$-matrix. On the other hand, one can Fourier transform the leading soft factor to position space to derive the  change in the asymptotic gauge or gravitational fields after the scattering process. This is known as the memory effect, and in particular one can derive the linear gravitational memory effect~\cite{BT:1987} from the Fourier transform of the leading soft graviton factor. The infrared triangle is conjectured to hold for the soft expansion with appropriate modifications. For instance, the Fourier transform of the subleading $\ln \omega^{-1}$ term in \ref{sf.gen} provides a subleading tail to the leading memory effect \cite{Laddha:2018myi, Fernandes:2020tsq}. 

The second development in recent years involves a novel proposal for flat spacetime holography, based on a celestial conformal field theory (CCFT) defined on the celestial sphere at the asymptotic boundary of flat spacetime \cite{Raclariu:2021zjz, Pasterski:2021rjz}. This conformal theory can be derived by transforming the usual momentum eigenbasis to boost scattering states. Hence asymptotic plane waves map to celestial conformal primary wavefunctions on the celestial sphere at null infinity, while scattering amplitudes map to celestial correlation functions. These correlators satisfy the symmetries of scattering amplitudes in flat spacetimes, and as a consequence, they have several differences with correlation functions of a standard conformal field theory (CFT) in two spacetime dimensions. Some of these differences include conformal primary wavefunctions with a complex spectrum (having a scaling dimension $\Delta = 1 + i \lambda$  with $\lambda \in \mathbb{R}$) providing a complete basis for normalizable massless scattering states, and correlation functions with a $\delta(z-\bar{z})$ contribution, with $z$ the conformal cross ratio~\cite{Pasterski:2016qvg,Pasterski:2017kqt,Atanasov:2021cje,Arkani-Hamed:2020gyp}. The latter is the manifestation in celestial correlation functions of translation invariance in momentum space scattering amplitudes. Soft theorems also have an important role in that they constrain the operator product expansion of celestial correlators and hence the infrared structure of CCFTs~\cite{McLoughlin:2022ljp}. 

In light of the above, it is interesting to consider the existence of similar structures from the infrared limit of scattering processes on Anti-de Sitter (AdS) spacetimes,  as it provides us with our best understood example of holography -- the AdS/CFT correspondence \cite{Maldacena:1997re}. Some immediate problems arise on trying to explore generalizations of soft theorems more generally on backgrounds with a cosmological constant. For one, the relevant asymptotic observables are not scattering amplitudes due to the absence of asymptotically free states. In addition, the presence of an inherent length scale of the spacetime implies the absence of a strict soft limit as for particles on asymptotically flat spacetimes. On such spacetimes, we can however always consider the vanishing cosmological constant limit, or equivalently the large spacetime radius limit, to recover locally asymptotically flat regions on which scattering processes are governed by a $S$-matrix. We will be interested in asymptotically AdS spacetimes on which holographic aspects can be cleanly investigated.  In the large AdS radius ($L$) limit, the holographic dual of boundary correlation functions have been shown to reproduce the $S$-matrix in a flat spacetime region within AdS~\cite{Giddings:1999jq,Gary:2009ae,Gary:2009mi,Penedones:2010ue,Fitzpatrick:2011jn,Fitzpatrick:2011ia,Hijano:2019qmi}. More recently, the soft photon theorem for scattering amplitudes on this patch has also been realized as the $U(1)$ Ward identity satisfied by boundary operators using the AdS$_4$/CFT$_3$ correspondence~\cite{Hijano:2020szl}, where $\omega$ approaches zero, with the condition that $L$ goes to infinity first. Thus a road map to the broader goal can be posed through the following question -- Are there $L^{-1}$ corrections of flat spacetime results to leading order in frequency, that can provide a perturbative approach to the factorization of scattering in AdS?

In this paper, we review results derived in~\cite{Banerjee:2021llh} and \cite{Banerjee:2022oll} that confirm the existence of $L^{-2}$ corrected soft factors for the $S$-matrix in the $L\to \infty$ limit of AdS$_4$. The results in~\cite{Banerjee:2021llh} make use of classical soft theorems~\cite{Laddha:2018myi,Laddha:2018rle,Laddha:2018vbn,Laddha:2019yaj,Saha:2019tub}, wherein the classical limits of soft factors can be derived from radiative solutions in a classical scattering processes. On asymptotically flat spacetimes, this approach is applicable when the soft particle wavelengths are far larger than the scattering impact parameter while the energy of the soft radiation is far less than the energy of the scatterer. The analysis in~\cite{Banerjee:2021llh} provides two additional aspects on $D \ge 4$ AdS spacetimes. The first of these is that the impact parameter, while large, has an upper bound of the AdS radius $L$. The second is that the `soft limit' of the radiation now involves a double scaling limit, in which we take $\omega \to 0$ as $L \to \infty$ while $\gamma = \displaystyle{\lim_{\omega \to 0; L \to \infty}} \omega L \gg 1$ is a large constant.
 This formally results in corrections of $\gamma^{-2}$ for each of the terms appearing in \ref{sf.gen}. We derived the leading $\gamma^{-2}$ corrections to the leading and first subleading terms in \ref{sf.gen} in the soft photon and soft graviton cases.

In \cite{Banerjee:2022oll}, we further established that the $L^{-2}$ corrected soft photon factor is that for the usual $S$-matrix in the flat spacetime patch within AdS. Our analysis followed \cite{Hijano:2020szl}, where the relationship between bulk gauge fields and the boundary $U(1)$ current was derived using the Hamilton, Kabat, Lifschytz and Lowe (HKLL) bulk reconstruction~\cite{Hamilton:2006az}. This approach furnishes a derivation of the soft factor, and realizes soft theorems as a $U(1)$ Ward identity with a fixed boundary current. On generalizing this approach to order $L^{-2}$, we can recover the classical soft theorem result. Yet another consequence from the analysis in \cite{Banerjee:2022oll} is that the HKLL kernels can be shown to admit an infinite series expansion in powers of $L^{-2}$ and thus our result provides the leading AdS radius correction to the soft photon factor.

We review the derivations and results of \cite{Banerjee:2021llh} and \cite{Banerjee:2022oll} in Sec.~\ref{sec2} and Sec.~\ref{sec3} respectively. Collectively, these results establish the existence of $L^{-2}$ corrections to the soft photon factor for scattering amplitudes in flat spacetime regions within AdS$_4$, and that they have a holographic interpretation. Directions for further research in light of these results are discussed in Sec.~\ref{sec4}

\section{Derivation from classical soft theorems} \label{sec2}

In this section, we briefly review the derivation of $L^{-2}$ corrected soft factors for flat spacetime regions within AdS$_4$ using classical soft theorems~\cite{Banerjee:2021llh,Banerjee:2020dww}.
We further discuss how the soft photon result motivates an equivalence between a perturbed large gauge Ward identity and the $L^{-2}$ corrected soft photon theorem. 

Due to the absence of asymptotically free states, AdS spacetimes lack a globally defined $S$-matrix. Hence the derivation of soft factors for scattering amplitudes in an infrared limit of the particles as on asymptotically flat spacetimes is not immediately clear. However, classical soft theorems~\cite{Laddha:2018myi,Laddha:2018rle,Laddha:2018vbn,Laddha:2019yaj,Saha:2019tub} allow for the derivation of classical contributions to soft photon and graviton factors for certain scattering processes. These theorems can be applied when the change in the energy of the scatterer does not significantly change over the process ($\Delta E_{\text{scatterer}}<<1$), while the emitted radiation wavelength is far larger than the impact parameter ($\lambda_{\text{radiation}} >> b$). Probe scattering processes are among those that realize these criteria. Such scattering processes along with the other essential requirement that they be gauge and diffeomorphism invariance, can be realized on AdS spacetimes. This motivates our analysis in this section on a globally AdS$_4$ spacetime.

For the electromagnetic case, classical soft theorems provide the following relationship between the classical limit of the soft photon factor and the emitted electromagnetic radiation of the scattering process
 
\begin{align}
\lim_{\omega \to 0}\epsilon^{\mu} a_{\mu}\left(\omega\,, \vec{x}\right) &= e^{i \omega R_{\rm obs}} \left(\frac{\omega}{2 \pi i R_{\rm obs}}\right)^{\frac{D-2}{2}} \frac{1}{2 \omega} S^{\text{flat}}_{\text{em}} 
\label{Eq:softem}
\end{align}

In \ref{Eq:softem}, for a $D \ge 4$ dimensional spacetime $\epsilon^{\alpha}$ is an arbitrary polarization vector of the photon,  $a_{\mu}$ is the radiative solution for the electromagnetic field in frequency space, $S_{\rm em}^{\rm flat}$ is the soft photon factor and $R_{\rm obs}$ is the distance of soft photon from the scattering center.  

On spacetimes that possess a flat spacetime region, we can consider \ref{Eq:softem} to hold perturbatively. Hence a perturbative correction for a radiative solution $a_{\mu}$ on AdS spacetimes should provide an appropriate correction to the classical limit of soft factors of flat spacetimes. To investigate this further, we consider the Reissner-Nordstr\"om AdS$_4$ spacetime  in isotropic coordinates $(t\,,x^i)$. In these coordinates we have the metric components

\begin{align}
g_{00} = -\left(1 + 2 \phi+\frac{\rho^2}{L^2}\right) \,, \qquad g_{0i} = 0 \,, \qquad g_{ij} = \delta_{ij} \left(1 - 2\phi+\frac{\rho^2}{2L^2}\right)\,
\label{rn.meta}.
\end{align}

and gauge potential 

\begin{equation}
A_0(\vec{x}) = \frac{Q}{\rho}= -\frac{8 \pi Q}{M}\phi\left(\vec{x}\right),  \label{rn.gfa}
\end{equation}

where we defined

\begin{equation}
\phi\left(\vec{x}\right)= -\frac{M}{8 \pi \rho}\,.
\label{rnsol.pot}
\end{equation}

and $\rho = \vert \vec{x} \vert$. The scattering of a charged probe particle on this background can be carried out in a manner satisfying the criteria for classical soft theorems, as detailed in~\cite{Banerjee:2021llh,Banerjee:2020dww}. To this end, we introduce a probe with mass `$m$' and charge `$q$' moving along a worldline trajectory $r(\sigma)$, \cite{DeWitt:1960fc}

\begin{equation}
S_{P} = -m \int d\sigma \sqrt{-g_{\mu\nu}\frac{dr^{\mu}}{d \sigma}\frac{dr^{\nu}}{d \sigma}} + \frac{q}{4\pi} \int d\sigma A_{\mu}\frac{dr^{\mu}}{d \sigma}\,.
\label{action.pp}
\end{equation}

and consider linear perturbations of the spacetime. In \ref{action.pp} $\frac{dr^{\mu}}{d \sigma}=u^{\mu}$ is the tangent to the worldline of the probe. We computed the perturbations $h_{\mu\nu}$ of the metric and $a_{\mu}$ of the gauge field by respectively solving the Einstein and Maxwell equations. The procedure is based on the well known Synge world function formalism \cite{Peters:1966, Peters:1970mx, Kovacs:1977uw, Poisson:2011nh}. While this procedure is straightforward, the technical details are quite involved and we refer the interested reader to~\cite{Banerjee:2021llh,Banerjee:2020dww} for further details. In the following, we summarize the main consequences due to the presence of a large AdS radius $L$ and the result for the leading soft photon factor.

Probe scattering in the flat spacetime limit ($L \to \infty$) would require that we consider a large impact parameter, with $\rho \gg  G M \ge \sqrt{G} Q$ and hence where the black hole effectively acts as a point particle scatterer. In considering $L^{-1}$ corrections to this result, the regime for probe particle scattering gets modified to $GM << \rho << L$. As a consequence, perturbative solutions about the flat spacetime region appear in terms of dimensionless quantities constructed from $L$. This can affect the emitted electromagnetic and gravitational radiation and the trajectories of the hard probe particle, which we discuss in turn.  

Corrections to the flat spacetime radiative solutions are found to involve powers of $\omega^{-2} L^{-2}$. Thus a soft limit ($\omega \to 0$) for the perturbative solutions is well defined through a double scaling limit, where $\omega$ approaches zero as $L$ approaches infinity, while maintaining the product $\omega L \to \gamma$ as a large constant. The application of the double scaling limit on radiative solutions thus identifies corrections in powers of $\gamma^{-2}$ for soft factors in the flat spacetime region. We also note that frequency modes in global AdS are formally discrete and a continuous frequency spectrum in flat spacetime regions within AdS requires a double scaling limit. Here, this condition is required to make the perturbative approach about the flat spacetime region consistent with the soft limit of the emitted radiation.

For the probe particle, we can parametrize its asymptotic trajectory as

\begin{equation}
\vec{r}(t)=\vec{\beta}_{\pm}t -C_{\pm}\,\vec{\beta}_{\pm}\,\ln\vert t\vert+ D_{\pm}\frac{t^2}{L^2} \,,\label{posv}
\end{equation}

where $\vec{\beta}_{\pm}$ is a constant vector, $C_{\pm}$ and $D_{\pm}$ are constants, and $+$ ($-$) denotes the outgoing (incoming) probe particle. The $\ln\vert t\vert$ term is due to long range Coulombic interactions in $D=4$ spacetime dimensions, while the $\frac{t^2}{L^2}$ term results from the AdS potential. However the AdS potential term only contributes to radiative solutions to subleading $\mathcal{O}(L^{-4})$ corrections to the soft factor. This implies that the $\gamma^{-2}$ corrected soft factors should be relevant for the usual $S$-matrix in an asymptotically flat spacetime region within global AdS spacetimes.  

Lastly, we can apply \ref{Eq:softem} to derive soft factors by considering the low frequency expansion on the radiative solutions. We derived the electromagnetic and gravitational radiative solutions up to the leading ($\omega^{-1}$) and first subleading ($\ln \omega^{-1}$) terms, each being accompanied by their respective $\gamma^{-2}$ corrections. These terms provide corresponding expressions for soft factors using \ref{Eq:softem}. While this result is formally in terms of one incoming and one outgoing (probe) particle, it is straightforward to generalize the final result to multiple hard particles. In the following, we consider the case of the leading soft photon factor which is relevant for subsequent discussions. For a process involving $n$ hard particles with charges $Q_{(a)}$, momenta $p_{(a)}^{\mu}$ with $\eta_{(a)} = 1(-1)$ for outgoing (ingoing) hard particles, and a single soft photon with momentum $q^{\mu}$ and polarization $\epsilon_{\mu}$, we find the leading soft photon factor result
\begin{align}
     S^{(0)}_{\text{em}} &= S^{(0);\text{F}}_{\text{em}} + S^{(0);\text{L}}_{\text{em}} \label{lst}\\
    S^{(0);\text{F}}_{\text{em}} &= \sum_{a=1}^n  Q_{(a)} \eta_{(a)} \frac{\epsilon_{\mu} p^{\mu}_{(a)}}{p_{(a)}.q} \,, \label{fl.sf}\\ 
    S^{(0);\text{L}}_{\text{em}} &= \frac{\omega^2}{4 \gamma^2} \sum_{a=1}^n Q_{(a)} \eta_{(a)} \frac{\epsilon_{\mu} p^{\mu}_{(a)}}{p_{(a)}.q} \frac{\vec{p}^2_{(a)}}{\big(p_{(a)}.q\big)^2} \,,
\label{ads.sf}
\end{align}

\ref{lst} provides the leading $\omega^{-1}$ divergence in the soft photon factor, while the $S^{(0);\text{F}}_{\text{em}}$ and $S^{(0);\text{L}}_{\text{em}}$ respectively denote the flat spacetime result and its $\gamma^{-2}$ correction. 

While the above result is for the soft photon factor, it can be used to predict a perturbed soft photon theorem across null infinity of the flat spacetime region within AdS$_4$. This follows from two observations, one for the leading soft factor and the other on the corresponding scattering process. The leading soft factor (in both gauge and gravitational theories) is universal for scattering amplitudes to all loop orders. We in addition noted that the soft factor up to $L^{-2}$ corrections are those for scattering processes whose asymptotic trajectories are those for asymptotically flat spacetimes. Hence the soft factor in \ref{lst} can be considered as that for a hard scattering process described by a $S$-matrix in the flat spacetime region within AdS$_4$. It also follows we require a perturbed soft photon mode to satisfy a soft theorem, with \ref{lst} the corresponding soft factor for the amplitude. Denoting the scattering amplitude for a hard process within the flat spacetime region as $\mathcal{S}$, we have a perturbed soft photon theorem  

 \begin{align}
 \lim_{\omega \rightarrow 0} \omega &\left(\langle \text{out}\vert \hat{a}_{\vec{q}}^{\text{out; F}}\mathcal{S} \vert \text{in} \rangle + n_L \langle \text{out}\vert \hat{a}_{\vec{q}}^{\text{out; L}\,(+)}(\omega \hat{x})\mathcal{S}\vert \text{in} \rangle\right) = \left(S^{(0);\text{F}}_{\text{em}} + S^{(0);\text{L}}_{\text{em}}\right) \langle \text{out}\vert \mathcal{S} \vert \text{in} \rangle
 \label{sptaa.corr}
 \end{align}

with $\hat{a}_{\vec{q}}^{\text{out;F}(+)}$ the flat spacetime soft photon mode that provides the Weinberg soft photon factor $S^{(0);\text{F}}_{\text{em}}$, while $\hat{a}_{\vec{q}}^{\text{out; L}\,(+)}$ is a perturbed mode responsible for the $\gamma^{-2}$ corrected soft factor $S^{(0);\text{L}}_{\text{em}}$. The form of the Ward identity in \ref{sptaa.corr}, in referring only to the outgoing positive helicity soft modes, follows from the $CPT$ invariance of the amplitude~\cite{Strominger:2017zoo}. We also note that while classical soft theorems provide results for soft factors, they provide no inference on the $S$-matrix or the soft modes. Thus \ref{sptaa.corr} involves a constant $n_L$ as the coefficient of the perturbed mode as a possible ambiguity.

We will now evaluate \ref{sptaa.corr} for a massless scattering process, which will be relevant for comparison with the results of the next section. We assume $k$ incoming and outgoing massless hard particles with momenta $p^{\mu}_k$, energies $E_{k} > 0$ and coordinates $\{z_k \,, \bar{z}_k\}$ for the particles location on null infinity of the flat region, and a single outgoing photon with coordinates $\{z_q \,, \bar{z}_q\}$, energy $\omega$ (which will be taken to be soft) and polarization $\epsilon^{\mu}_+$

\begin{align}
p^{\mu}_k & = \eta_k \frac{E_k}{1 + z_k\bar{z}_k}\left(1+z_k\bar{z}_k\,, z_k + \bar{z}_k\,, -i (z_k - \bar{z}_k)\,, 1 - z_k\bar{z}_k\right) \,, \notag\\
q^{\mu} & = \frac{\omega}{1 + z_q\bar{z}_q}\left(1+z_q\bar{z}_q\,, z_q+ \bar{z}_q\,, -i (z_q - \bar{z}_q)\,, 1 - z_q\bar{z}_q\right) \notag\\
\epsilon^{\mu}_+ &= \frac{1}{\sqrt{2}} \left(\bar{z}_q\,, 1 \,, -i \,, -\bar{z}_q\right)\,,
\label{p.hs}
\end{align}

We can use \ref{p.hs} to derive expressions for 
 $S^{(0);\text{F}}_{\text{em}}$ and $S^{(0);\text{L}}_{\text{em}}$ in \ref{lst}, and subsequently \ref{sptaa.corr}. For the flat spacetime mode, we recover the Weinberg soft theorem result

\begin{align}
& \lim_{\omega \rightarrow 0} \frac{\sqrt{2} \omega}{\left(1 + z_q \bar{z}_q\right)}\langle \text{out}\vert \hat{a}_{\vec{q}}^{\text{out; F}\, (+)}(\omega \hat{x}) \mathcal{S} \vert \text{in}\rangle =  \sum_{k} \eta_k Q_k \frac{1}{z_q-z_k}\langle \text{out}\vert \mathcal{S} \vert \text{in} \rangle\,,
\label{wst}
\end{align}

while the corrected mode satisfies

\begin{align}
& \lim_{\omega \rightarrow 0} \frac{\sqrt{2} \omega n_L}{\left(1 + z_q \bar{z}_q\right)}\langle \text{out}\vert \hat{a}_{\vec{q}}^{\text{out; L}\, (+)}(\omega \hat{x}) \mathcal{S} \vert \text{in} \rangle =  \frac{1}{16 \gamma ^2} \sum_{k} \eta_k Q_{k} \frac{\left(1 + z_k \bar{z}_k\right)^2 \left(1 + z_q \bar{z}_q\right)^2}{(\bar{z}_q-\bar{z}_k)^2 (z_q-z_k)^3} \langle \text{out}\vert \mathcal{S} \vert \text{in} \rangle
\label{lpt1}
\end{align}

Thus classical soft theorems predict a $\gamma^{-2}$ correction of the Weinberg soft photon theorem in a flat region within AdS$_4$ spacetimes that arises in taking $L \to \infty$. While the scattering process in the flat region may involve corrections to the $S$-matrix, the $\gamma^{-2}$ corrections of the soft factor are those for a flat spacetime $S$-matrix (involving hard particles with asymptotic trajectories of a flat spacetime). The appearance of $L^{-2}$ corrections as $\gamma^{-2}$ arise from a double scaling limit (on the frequency and AdS radius) as the appropriate soft limit on AdS spacetimes. 

A remarkable property of \ref{wst} is that it can also be derived from large gauge transformations and is equivalent to a large gauge Ward identity~\cite{Strominger:2017zoo}. This follows from gauge transformations on the asymptotic fields. In this regard, we note that \ref{lpt1} may likewise be realized from a perturbed Ward identity resulting from $L^{-2}$ corrections of the gauge parameter for large gauge transformations ~\cite{Banerjee:2021llh}. In the following section, we will be interested in an AdS$_4$ specific correspondence -- the equivalence of \ref{sptaa.corr} with a $U(1)$ Ward identity of a boundary CFT$_3$.

\section{Derivation from CFT$_3$ Ward identity} \label{sec3}

We will now consider the derivation of $L^{-2}$ corrections to the known soft photon theorem for a $S$-matrix defined on the asymptotically flat spacetime patch in AdS$_4$. Our approach follows that of \cite{Hijano:2020szl}, generalized to $L^{-2}$ corrections. We consider the AdS$_4$ spacetime metric in global coordinates
\begin{equation}
ds^2 = \frac{L^2}{\cos^2(\rho)}\left[- d\tau^2 + d \rho^2 + \sin^2(\rho) d\Omega_2^2\right]\,,
\label{ads.gm}
\end{equation}
with $d\Omega_2^2$ in terms of complex stereographic coordinates $\{z\,,\bar{z}\}$. Substituting 

\begin{equation}
\rho = \arctan \left(\frac{r}{L}\right) \;; \qquad  \tau = \frac{t}{L}  \label{fc}
\end{equation}
in \ref{ads.gm} and taking $L \to \infty$, we recover an asymptotically flat spacetime patch centered around global time $\tau = 0$. We will be interested in the description of a scattering process in this region. From the classical soft theorem result, we further expect corrections to soft factors of the $S$-matrix in the flat spacetime patch. In the previous section, we noted that this correction can be associated with a perturbed soft photon mode. This will be further established from the analysis in this section. One possible interpretation of the result is that the soft photon can propagate beyond the flat spacetime region, which manifests as a perturbed mode on the boundary of the flat spacetime region.

In the flat spacetime region, we adopt coordinates $y = \{t\,, \vec{y}\} =  \{t\,, r\,,z\,,\bar{z}\}$. The photon momentum is parametrized as  $q = \omega \{1 \,, \hat{q}\}$ with $\omega$ the frequency, $\hat{q}$ the orientation and $q^2 = 0$. The spatial component is thus $\vec{q} = \omega \hat{q}$. The photon polarization vectors are denoted by $\varepsilon^{(\lambda)}_{\mu}$ which are normalized as $\varepsilon^{(+)}_{\mu}\varepsilon^{(-) \mu} = 1$ and where $\lambda$ is the helicity ($\pm$) . We may then define photon creation and annihilation modes from gauge fields in position space for the flat spacetime region in the usual way

\begin{align}
 \hat{a}_{\vec{q}}^{(\lambda)} &= \lim_{t\rightarrow\pm\infty} i \int d^3 \vec{y} \,  \varepsilon^{(\lambda) \mu}  e^{-i q\cdot y} \overleftrightarrow{\partial_0} \hat{{\cal A}}_{\mu}(y) \,, \label{a.f} \\
\hat{a}_{\vec{q}}^{(\lambda) \dagger} &= \lim_{t\rightarrow\pm\infty} - i \int d^3 \vec{y} \,  \varepsilon^{(\lambda) * \mu}  e^{i q\cdot y} \overleftrightarrow{\partial_0} \hat{{\cal A}}_{\mu}(y) \,,\label{ad.f}
\end{align}

with $t \to \infty$ ($t \to - \infty$) providing the outgoing (ingoing) modes in \ref{a.f} and \ref{ad.f}. These modes satisfy the canonical commutation relations

\begin{equation}
\left[\hat{a}_{\vec{q}}^{(\lambda)} \,,  \hat{a}_{\vec{q}\,'}^{(\lambda')}\right] = \delta^{\lambda \lambda'} (2 \pi)^3 2\omega_{q} \delta^{(3)}(\vec{q} - \vec{q}\,')\,.
\label{cc}
\end{equation}

 In addition, the choice of spherical coordinates in \ref{ads.gm} leads to the following expressions for the positive helicity polarization vector and outgoing plane wave state

\begin{align}
\varepsilon^{(+) z}  =  \frac{1 + z \bar{z}}{\sqrt{2} r}, \quad 
 e^{-i q\cdot y}  = e^{i \omega t} e^{-i \vec{q}.\vec{r}} = e^{i \omega t} 4\pi \sum_{l',m'} (-i)^{l'} j_{l'}(r \omega) Y_{l'm'}(\Omega) Y^*_{l'm'}(\Omega_{q})\,,
\label{pl}
\end{align}

with $Y_{lm}(\Omega)$ the spherical harmonics and $j_{l}(v)$ the spherical Bessel function of order $l$ that is related to the Bessel function $J_{k}(v)$ via

\begin{equation}
j_l(v) = \sqrt{\frac{\pi}{2 v}} J_{l+\frac{1}{2}}(v)\,.
\label{sb}
\end{equation}

Similarly for the hard particles involved in the scattering, we can define flat spacetime modes in terms of bulk fields in position space. We will be interested in applying a Lorentzian HKLL bulk reconstruction to express \ref{a.f} and \ref{ad.f} in terms of derivatives of a boundary $U(1)$ current.  This will follow from relating $\hat{{\cal A}}_{\mu}(y)$ with the boundary current using the HKLL reconstruction, which can be expanded in powers of $L^{-2}$. As a consequence, we may derive the $L^{-2}$ corrected soft photon modes using \ref{a.f} and \ref{ad.f} from the corresponding $L^{-2}$ corrected expressions for $\hat{{\cal A}}_{\mu}(y)$ in terms of the boundary current. 

There are additional aspects regarding the reconstruction of flat spacetime patch modes from boundary operators of AdS$_4$, which identify a correspondence between the boundary regions~\cite{Hijano:2020szl}. We now briefly recall the general procedure for all flat spacetime modes before turning our attention to the derivation of soft modes. The AdS$_4$ spacetime is foliated in terms of global Cauchy slices at constant $\tau$, whose $L \to \infty$ limit recover asymptotic slices of the flat spacetime region. Formally, the domain of dependence is $\tau \in \{\epsilon\,, \frac{\pi}{2} - \epsilon\}$ and $\tau \in \{-\frac{\pi}{2} + \epsilon\,, -\epsilon \}$ respectively for past and future Cauchy slices. The scattering process in the flat spacetime region takes place over a $\mathcal{O}(L^{-1})$ window about $\tau = 0$. The bulk fields in position space can be expressed in terms boundary CFT currents using HKLL kernels~\cite{Hamilton:2006az} , which provide a way to derive modes of the flat spacetime region in terms of CFT boundary operators. The procedure when applied to massless and massive bulk fields has a $\tau$ dependent phase that identifies a correspondence between asymptotic regions of the flat spacetime patch and regions of the AdS boundary in the $L \to \infty$ limit. Future (past) bulk massless fields are reconstructed from boundary operators in a $\mathcal{O}(L^{-1})$ window about $\tau = \frac{\pi}{2}$ ($\tau = -\frac{\pi}{2}$) denoted by $\tilde{\mathcal{I}}^+$ ($\tilde{\mathcal{I}}^-$). Future (past) bulk massive fields on the other hand have a complex $\tau$ dependence and in the $L \to \infty$ limit are reconstructed from analytically continued Euclidean regions at $\tau = \frac{\pi}{2}$ ($\tau = -\frac{\pi}{2}$) denoted as $\partial \mathcal{M}_+$ ($\partial \mathcal{M}_-$). In this way null, timelike and spacelike infinity of the flat spacetime region can be identified with corresponding regions of the AdS boundary, as depicted in Figure 1.

\begin{figure}[h]
    \begin{center}
        \includegraphics[scale = 0.3]{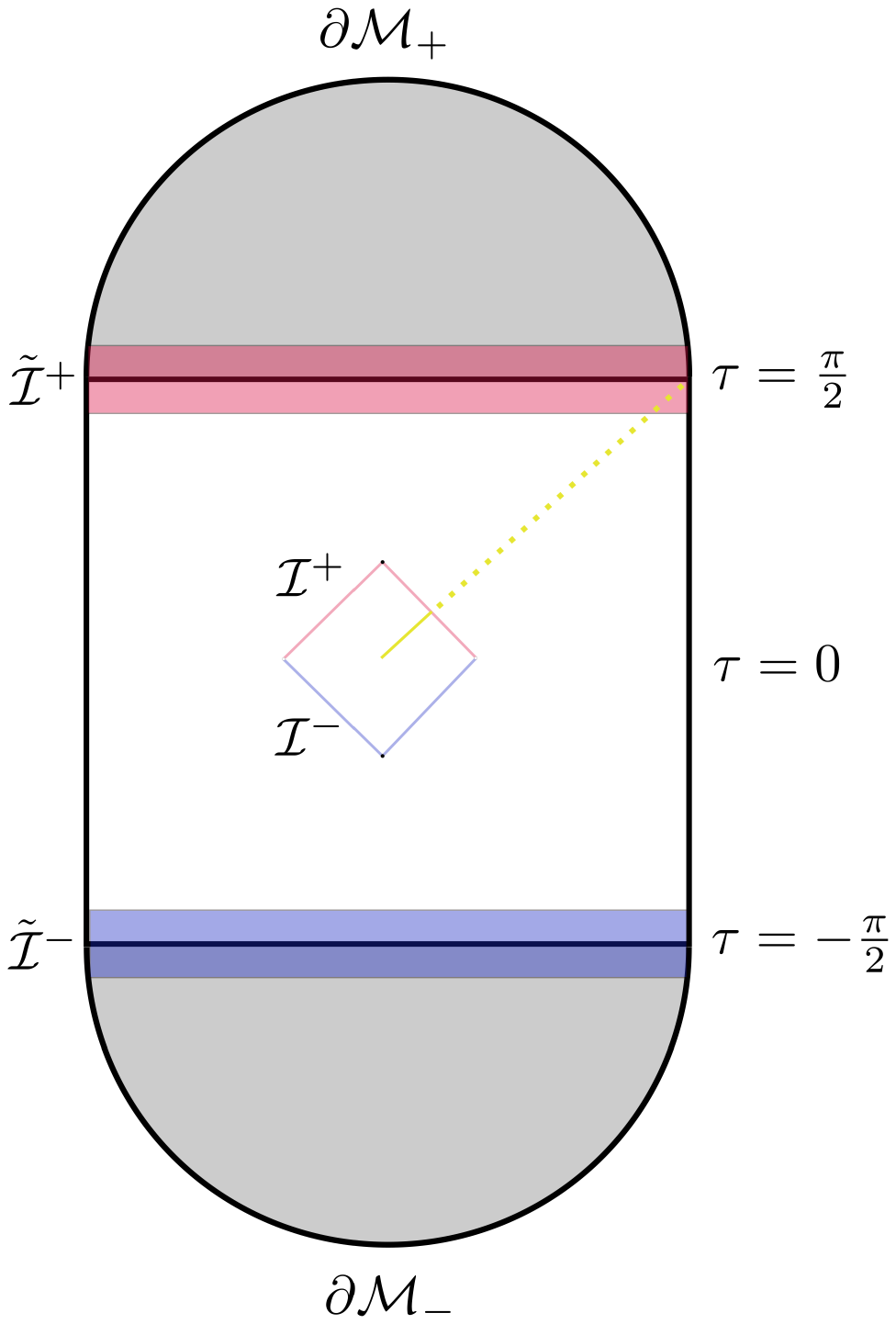}
    \end{center}
    \caption{Massless bulk field modes defined across future (pink) and past (blue) null infinities of the flat spacetime patch have a respective correspondence with boundary currents on $\tilde{\mathcal{I}}^+$ and $\tilde{\mathcal{I}}^-$ of the AdS boundary. This support is identified from the $L \to \infty$ limit of bulk fields which define modes on the flat spacetime patch. The $\tilde{\mathcal{I}}^{\pm}$ are defined around $\tau' = \pm \frac{\pi}{2}$. We have further indicated the trajectory of a soft particle (yellow), whose global propagation (dashed) might be considered as manifesting in $\gamma^{-2}$ corrected soft factors on the flat patch boundary. A similar analysis on massive fields provides a correspondence between timelike infinities $i^{\pm}$ of the flat patch and Euclidean caps $\partial \mathcal{M}_{\pm}$, while spatial infinity $i_0$ is related with the $\tau \in \{-\frac{\pi}{2}\,,\frac{\pi}{2}\}$ region.}
\end{figure}

The bulk gauge field in AdS$_4$ can be reconstructed from a CFT$_3$ current by specifying an asymptotic boundary condition and describing the modes in terms of boundary integrated expressions over HKLL kernels. We denote a bulk point in AdS$_4$ by $X = \{\tau \,, \rho \,, z\,,\bar{z}\}$ and a boundary point by $x' = \{\tau'\,,z'\,,\bar{z}'\}$. The scattering process we will be interested in involve massless hard particles, and hence have no Coulombic modes. Hence on a fixed $\tau$ Cauchy slice, we impose the following holographic boundary condition

\begin{equation}
\lim_{\rho \to \frac{\pi}{2}}\hat{\mathcal{A}}_{\mu} (\rho\,,x') = \cos (\rho) j_{\mu}(x')\,,
\label{A.bc}
\end{equation}

with $j_{\mu}(x')$ a conserved $U(1)$ boundary current of conformal dimension $\Delta = 2$, acting as the source for the bulk gauge field. 
The choice \ref{A.bc} will provide the Weinberg soft photon theorem in the absence of magnetic charges. Due to the additional absence of Coulombic fields, the bulk gauge field is radiative with non-vanishing components for $\mu=z\,, \bar{z}$.

The integrated expressions over HKLL kernels follow from solutions of Maxwell's equations in AdS$_4$ considered in a spherical harmonic decomposition~\cite{Ishibashi:2004wx}. There exist `vector' ($\Delta =2$) and `scalar' ($\Delta =1$) type solutions, which respectively will be in terms of the curl and divergence of the boundary current. The HKLL presription then provides the following expression 

\begin{align}
\hat{{\cal A}}_{\mu}(X) &= \int d^3 x' \left[K^V_{\mu} (X\,;x') \epsilon_{\tau'}^{a' b'} \nabla_{a'} j^+_{b'} + K^S_{\mu}(X \,;x') \gamma^{a' b'} \nabla_{a'} j^+_{b'} \right.  \notag\\
&  + \left. \left(K^V_{\mu}\right)^*(X \,;x') \epsilon_{\tau'}^{a' b'} \nabla_{a'} j^-_{b'} + \left(K^S_{\mu}\right)^*(X \,;x') \gamma^{a' b'} \nabla_{a'} j^-_{b'} \right]\,,
\label{mk}
\end{align}

where $\epsilon^{a' b' c'}$ and $\nabla_{a'}$ respectively are the boundary Levi-Civita tensor and covariant derivative, while the ${\pm}$ signs on $j^{\pm}_{b'}$ indicate positive and negative frequency solutions of the boundary current. The boundary integral is
\begin{equation} 
\int d^3 x' = \int\limits_{\mathcal{T}} d\tau' \int d \Omega'  
\label{bi}
\end{equation}

with the $\tau'$ integral being $\{-\pi\,,0\}$ for ingoing states and $\{0\,,\pi\}$ for outgoing states, simply denoted as $\mathcal{T}$ in \ref{bi}. The $K^V_{\mu}$ and $K^S_{\mu}$ in \ref{mk} are respectively HKLL kernels for `vector' and `scalar' type components of the Maxwell field, with non-vanishing components 

\begin{align}
K^V_{z} (X\,;x') &= \frac{1}{\pi} \sum_{\kappa\,, l\,, m} \mathcal{N}^V  Y^*_{lm}\left(\Omega'\right) \partial_{z}Y_{lm}\left(\Omega \right) \Xi_{\kappa l}(\rho\,,\tau\,,\tau')\Big\vert_{\Delta =2} \label{kv1}\\
K^V_{\bar{z}} (X\,;x') &= - \frac{1}{\pi} \sum_{\kappa\,, l\,, m} \mathcal{N}^V  Y^*_{lm}\left(\Omega'\right) \partial_{\bar{z}}Y_{lm}\left(\Omega \right) \Xi_{\kappa l}(\rho\,,\tau\,,\tau')\Big\vert_{\Delta =2} \label{kv2}\\
K^S_{z} (X\,;x') &=  \frac{1}{\pi} \sum_{\kappa\,, l\,, m} \mathcal{N}^S Y^*_{lm}\left(\Omega'\right) \partial_{z}Y_{lm}\left(\Omega \right) \Xi_{\kappa l}(\rho\,,\tau\,,\tau')\Big\vert_{\Delta =1} \label{ks1}\\
K^S_{\bar{z}} (X\,;x') &=  \frac{1}{\pi} \sum_{\kappa\,, l\,, m} \mathcal{N}^S Y^*_{lm}\left(\Omega'\right) \partial_{\bar{z}}Y_{lm}\left(\Omega \right) \Xi_{\kappa l}(\rho\,,\tau\,,\tau')\Big\vert_{\Delta =1} \;, \label{ks2}
\end{align}
and
\begin{equation}
\Xi_{\kappa l}(\rho\,,\tau\,,\tau') = e^{i \omega_{\kappa} (\tau - \tau')} \sin^{l+1} \rho \cos^{\Delta-1} \rho \,_2F_1\left(- \kappa \,, \kappa + \Delta + l \,, \Delta - \frac{1}{2}\Big\vert \cos^2 \rho \right)\,,
\label{xi}
\end{equation}

By substituting $\hat{{\cal A}}_{\mu}(X)$ in \ref{a.f} and \ref{ad.f}, we can thus derive the flat spacetime and leading perturbed soft photon mode by expanding up to $L^{-2}$ corrections. Subleading AdS radius corrections in \ref{mk} can arise from $\Xi_{\kappa l}$ in \ref{xi} and possibly the normalizations in \ref{kv1} - \ref{ks2}. We first discuss the contribution from \ref{xi}. The frequency modes $\omega_{\kappa}$ in \ref{xi} can be noted as being discrete (running over positive integers) and has the following relation with the scaling dimension $\Delta$, discrete energy label $\kappa$ and angular momentum mode $l$ 

\begin{equation}
\omega_{\kappa} =  2 \kappa + \Delta + l \,.
\label{adsfr}
\end{equation}

These modes can provide a continuous frequency $\omega$ in the flat spacetime region by requiring that they are dominated by large values of $\kappa$ and scale with $L$ in the $L \to \infty$ limit, i.e. $\omega_{\kappa} \approx \omega L$.  Then the sum over $\kappa$ can be traded for an integral over $\omega$ in \ref{kv1} - \ref{ks2}

\begin{equation}
\sum_{\kappa} \to \frac{1}{2}\int d\omega L
\end{equation}

On using $\omega_{\kappa} = \omega L$ and $\tau$ from \ref{fc} in \ref{xi} we find

$$e^{i \omega L (\tau - \tau')} = e^{i \omega t} e^{-i \omega L \tau'} \,,$$

and hence there are no $L^{-2}$ corrections from the phase in \ref{xi}. However, the substitution of $\rho$ from \ref{fc} in \ref{xi} does lead to a result that can be systematically expanded in powers of $L^{-2}$. The expansions leading up to this result are considerably involved and make use of several identities and expansions for the Euler Gamma and hypergeometric functions~\cite{Banerjee:2022oll}. We find the result

\begin{align}
&\Xi_{\kappa l}(\rho\,,\tau\,,\tau')\Big \vert_{\Delta = 2} =  - (\pm i)^{- l} e^{i \omega t} e^{-i \omega L \left( \tau' \mp \frac{\pi}{2}\right)} \frac{r}{L}\Bigg\{j_{l}(r \omega) \left( 1 + \frac{1}{2 \omega^2 L^2}\left(\frac{l(l+1)}{2} - \frac{(r \omega)^2}{3}\right)\right) \notag\\ 
&   \qquad \qquad \qquad \qquad  - \frac{1}{2 \omega^2 L^2} \sqrt{\frac{\pi}{2 r \omega}} \left( \frac{l (l+1)}{2} + (r \omega)^2 \right) \frac{2 r\omega}{3} J'_{l+\frac{1}{2}}(r \omega) \Bigg\} + \mathcal{O}\left( \frac{1}{\omega^3 L^3}\right)\,,\label{xi2}\\
& \Xi_{\kappa l}(\rho\,,\tau\,,\tau')\Big \vert_{\Delta = 1} = - (\pm i)^{- l + 1}  e^{i \omega t} e^{-i \omega L \left( \tau' \mp \frac{\pi}{2}\right)} (\omega L) \frac{r}{L} \Bigg\{j_{l}(r \omega) \left( 1 + \frac{1}{2 \omega^2 L^2}\left(-\frac{l(l+1)}{2} - \frac{(r \omega)^2}{3} \right)\right) \notag\\ 
&  \qquad \qquad \qquad \qquad  - \frac{1}{2 \omega^2 L^2} \sqrt{\frac{\pi}{2 r \omega}} \left( \frac{l (l+1)}{2} + (r \omega)^2 \right) \frac{2 r\omega}{3} J'_{l+\frac{1}{2}}(r \omega)\Bigg\} + \mathcal{O}\left( \frac{1}{\omega^3 L^3}\right) \,,\label{xi1}
\end{align}

 with the primes on the Bessel functions in \ref{xi2} and \ref{xi1} denoting derivatives with respect to the argument. We note that the corrections appear in terms of $\omega^{-2} L^{-2}$ and this leads to a well defined soft limit $\omega \to 0$ in the flat spacetime region if we adopt a double scaling limit $\omega \to 0$ as $L \to \infty$. In addition, consistent with the requirement of being dominated by large $\kappa$ modes, $\omega L$ must be large and hence the $\omega^{-2} L^{-2}$ corrections to the flat spacetime result are perturbative.

 The other source of $L^{-2}$ corrections are the normalization constants $\mathcal{N}^{V}$ and $\mathcal{N}^{S}$ in the kernels \ref{kv1} - \ref{ks2}, which are derived from the Klein-Gordon inner product applied to the Maxwell field solutions

 \begin{equation}
\mathcal{N}^V =  -\frac{1}{4 l(l+1)} \qquad \;; \qquad \mathcal{N}^S =  -\frac{1}{4 l(l+1)} \frac{i}{\omega L} \left(1 + \frac{l(l+1)}{2 \omega^2 L^2}\right)\,,
\label{norm}
 \end{equation}

Thus by substituting \ref{norm}, \ref{xi2} and \ref{xi1} in \ref{kv1} - \ref{kv2}, we can find expressions for the bulk gauge field up to $L^{-2}$ corrections from \ref{mk}. We can substitute this bulk gauge field in \ref{a.f} and consider the $t \to \infty$ solution and $\omega \to 0$ limit to derive positive helicity and outgoing soft photon modes. The $L^0$ contribution of the bulk gauge field recovers the result~\cite{Hijano:2020szl}

\begin{equation}
\lim_{\omega \to 0} \omega  \frac{\sqrt{2}}{1+z_q \bar{z}_q} \hat{a}_{\vec{q}}^{\text{out; F}\,(+)} =  \frac{1}{4}  \int\limits_{\tilde{\cal{I}}^+} d\tau' \int d\Omega' \, 
 \frac{1}{z_q-z'} D^{\bar{z}'}j^-_{\bar{z}'}(x')\,,
 \label{af}
\end{equation}

while the $L^{-2}$ corrected bulk gauge field gives the corrected soft mode~\cite{Banerjee:2022oll}

\begin{equation}
\lim_{\omega \to 0} \omega  \frac{\sqrt{2}}{1+z_q \bar{z}_q} \hat{a}_{\vec{q}}^{\text{out; L}\, (+)} = \frac{1}{32 \pi \gamma^2} \int\limits_{\tilde{\cal{I}}^+} d\tau' \int d\Omega' \int d\Omega_{w} \, \left[\frac{\left(1+ z' \bar{z}'\right)^2\left(1+ z_w \bar{z}_w\right)^2}{\left(\bar{z}' - \bar{z}_{w}\right)^2 \left(z_q - z_w\right)^3}\right] \mathcal{D}^{\bar{z}'}j^{-}_{\bar{z}'} 
\label{aL}
\end{equation}

We briefly explain the the integral over $\tau'$ over $\tilde{\cal{I}}^+$ in the boundary of AdS$_4$. The results in \ref{af} and \ref{aL} formally involve a phase $e^{i \omega L \left( \tau' - \frac{\pi}{2}\right)}$, 
 which is highly oscillatory except in a $\mathcal{O}(L^{-1})$ window about $\tau' = \frac{\pi}{2}$. This results in a support over the region $\tilde{\cal{I}}^+$ in the boundary of AdS$_4$ as previously discussed. However, if the domain of the integrand is understood as being supported over $\tilde{\cal{I}}^+$, we may simply consider the integrals as in in \ref{bi} over the entire future boundary.  In addition, we have a $\omega^{-2} L^{-2} \to \gamma^{-2}$ in \ref{aL} upon taking the soft limit.

As demonstrated in~\cite{Hijano:2020szl}, the $U(1)$ Ward identity can be shown to be equivalent to the Weinberg soft photon theorem. We recall this in the following before addressing the derivation of the $L^{-2}$ perturbed soft photon theorem using \ref{aL}

The $U(1)$ Ward identity has the integrated expression

\begin{equation}
\int d^3 x'\,\alpha(x')  \partial'_{\mu} \langle 0 \vert  T\{   j^{\mu}(x')  \Phi  \} \vert  0 \rangle  = \left(\sum_{i=1}^n Q_i \alpha (x'_i)- \sum_{j=1}^m   Q_j \alpha (x'_j) \right) \langle 0 \vert T\{ \Phi \}  \vert  0 \rangle
\, ,
\label{cw}
\end{equation}

with $T\{\cdots \}$ denoting the time ordering of the operators inside the parenthesis, and $\Phi$ a collection of CFT operators comprising of $n$ `ingoing' operators ($ \tau < 0 $) with charges $Q_i$ and $m$ `outgoing' operators ($\tau > 0$) with charges $Q_j$. In particular, we recover the $S$-matrix in the flat spacetime region from $\langle 0 \vert T\{ \Phi \}  \vert  0 \rangle$ in the $L \to \infty$ limit. The parameter $\alpha(x')$ can be specified to have the boundary expression

\begin{equation} 
\alpha(x')\vert_{\tilde{\cal{I}}^+}= \epsilon(\hat{x}') = \frac{1}{z_q - z'}\,,
\label{gf.a}
\end{equation}

From \ref{gf.a} we can recover the Weinberg soft photon theorem \ref{wst}. The right hand side of \ref{wst} is manifestly recovered, while the left hand side follows from the relationship between soft photon modes and derivatives of the boundary current \ref{af}. Here we note that formally we need to consider the sum over all incoming and outgoing soft photon modes. However, upon using the $CPT$ invariance of matrix elements with incoming and outgoing soft photons, the left hand side of \ref{wst} results simply from \ref{af} with a factor of 4. In this way, the $U(1)$ Ward identity on $\tilde{\cal{I}}^+$ of the AdS$_4$ boundary recovers the Weinberg soft photon theorem on $\cal{I}^+$ of the flat spacetime patch in the $L \to \infty$ limit.

\subsection{Recovering the classical soft photon theorem result} \label{match}

The derivation of the $L^{-2}$ corrected soft photon theorem from the $U(1)$ Ward identity in \ref{cw} can likewise be considered. From the preceding analysis, we note the crucial requirement on the choice of gauge parameter $\epsilon(\hat{x}')$ in \ref{gf.a} to recover the Weinberg soft factor. The choice can be inferred directly from \ref{af} and \ref{aL} by re-expressing these results as

\begin{align}
\lim_{\omega \to 0} \omega  \frac{\sqrt{2}}{1+z_q \bar{z}_q} \hat{a}_{\vec{q}}^{\text{out; F}\,(+)}=&  \frac{1}{4} \int d^3 x' \epsilon(\hat{x}') D^{\bar{z}'}j^-_{\bar{z}'}(x')\label{gp.f}\\
\lim_{\omega \to 0} \omega  \frac{\sqrt{2}}{1+z_q \bar{z}_q} \hat{a}_{\vec{q}}^{\text{out; L}\,(+)}=&  \frac{1}{4} \int d^3 x' \epsilon^{\text{L}}(\hat{x}') D^{\bar{z}'}j^-_{\bar{z}'}(x') \label{gp.L}
\end{align}

with $\epsilon(\hat{x}')$  as in \ref{gf.a} and $\epsilon^{\text{L}}(\hat{x}')$ defined as

\begin{align}
\epsilon^{\text{L}}(\hat{x}') &= \frac{1}{8 \pi \gamma^2}\int d\Omega_{w} \left[\frac{\left(1+ z' \bar{z}'\right)^2\left(1+ z_w \bar{z}_w\right)^2}{\left(\bar{z}' - \bar{z}_{w}\right)^2 \left(z_q - z_w\right)^3}\right] 
\label{gf.l1}
\end{align}

Thus by now considering 
\begin{equation} 
\alpha(x')\vert_{\tilde{\cal{I}}^+}= \epsilon^{\text{L}}(\hat{x}') \,,
\label{gf.l}
\end{equation}

in \ref{cw} and proceeding exactly as in the derivation of the Weinberg soft photon theorem, we now arrive at the following $L^{-2}$ corrected soft photon theorem

\begin{equation}
\lim_{\omega \rightarrow 0} \frac{\sqrt{2} \omega }{\left(1 + z_q \bar{z}_q\right)}\langle \text{out}\vert \hat{a}_{\vec{q}}^{\text{out; L}\, (+)}(\omega_q \hat{x}) \mathcal{S} \vert \text{in} \rangle =  \left[\ \sum_{k = \text{out}} \epsilon^{\text{L}}(x') Q_{k}  - \sum_{k=\text{in}} \epsilon^{\text{L}}(x') Q_k \right] \langle \text{out}\vert \mathcal{S} \vert \text{in} \rangle \,,
\label{spt.l}
\end{equation}

The result in \ref{spt.l} however does not immediately agree with \ref{lpt1}, as the former involves an integration over intermediate angles that is absent in the latter. On inspection, the two expressions will agree in a collinear limit of \ref{spt.l}. In fact, the leading soft photon factor is both soft ($\omega \to 0$) and collinear ($z_q \to z'$) divergent~\cite{Strominger:2017zoo}, which is the case for \ref{lpt1} but not \ref{spt.l}. Hence the procedure we now discuss may be considered as a requirement to derive a leading soft photon theorem with expected properties.   

Due to the large $L$ approximation, we expect a small separation between points on $\cal{I}^+$ of the flat patch parametrized by $\{z_q\,,\bar{z}_q\}$ and those on $\tilde{\mathcal{I}}^+$ of the AdS$_4$ boundary parametrized by $\{z'\,,\bar{z}'\}$. We thus consider $\vert z_q - z' \vert \approx \tilde{\epsilon}$ with $\tilde{\epsilon} \ll 1$. In addition, we can consider taking the collinear limit of $\{z_w\,,\bar{z}_w\}$ with either $\{z_q\,,\bar{z}_q\}$ or $\{z'\,,\bar{z}'\}$ by considering the following expansion 

\begin{equation}
    z_w = z_q + \delta e^{i \theta} \,, \quad z_w = z' + \delta e^{i \theta}\,,
    \label{zwe}
\end{equation}
where we made use of the property that $\vert z_q - z' \vert \approx \tilde{\epsilon}$ to describe an equivalent expansion about $\{z_q\,,\bar{z}_q\}$ and $\{z'\,,\bar{z}'\}$. We can now use \ref{zwe} to express the integrand in \ref{gf.l1} in terms of a leading piece independent of $\delta$ and subleading corrections that are $\mathcal{O}(\delta)$. The collinear limit thus corresponds to taking $\delta \to 0$. The integration in \ref{gf.l1} can be carried out to find the following result

\begin{align}
    \epsilon^{\text{L}}(\hat{x}') &=\frac{1}{ 2 \gamma^2} \frac{(1+z_q\bar{z}_q)^2 (1+z'\bar{z}')^2}{(\bar{z}_q - \bar{z}')^2 (z_q - z')^3}
     + \text{corrections}\,,
     \label{gf.l2}
 \end{align}

With the corrections in \ref{gf.l2} indicating the contributions from the $\mathcal{O}(\delta)$ corrections that are subleading in the collinear limit being considered. Hence using \ref{gf.l2} in \ref{spt.l}, we find

\begin{align}
& \lim_{\omega_q \rightarrow 0} \frac{\sqrt{2} \omega_q }{\left(1 + z_q \bar{z}_q\right)}\langle \text{out}\vert \hat{a}_{\vec{q}}^{\text{out; L}\, (+)}(\omega_q \hat{x}) \mathcal{S} \vert \text{in} \rangle \notag\\
 & \qquad \approx  \frac{1}{2 \gamma ^2}\left[\ \sum_{k = \text{out}} \frac{\left(1 + z' \bar{z}'\right)^2 \left(1 + z_q \bar{z}_q\right)^2}{(\bar{z}_q-\bar{z}')^2 (z_q-z')^3}Q_{k}  - \sum_{k=\text{in}} \frac{\left(1 + z' \bar{z}'\right)^2 \left(1 + z_q \bar{z}_q\right)^2}{(\bar{z}_q -\bar{z}')^2 (z_q - z')^3}Q_k \right] \langle \text{out}\vert \mathcal{S} \vert \text{in} \rangle\notag\\
 &\qquad \qquad \qquad \qquad \qquad \qquad  + \text{corrections}
\label{spt.l2}
\end{align}

Thus we find that \ref{spt.l2} agrees with \ref{lpt1} in a collinear approximation with the identification of $n_L = \frac{1}{8}$. The derivation using the $U(1)$ Ward identity hence fixes the overall normalization of the soft photon mode that was undetermined from classical soft theorems.

\section{Discussion and Outlook} \label{sec4}

In this paper, we discussed the soft factorization for scattering processes that arise in the large AdS radius limit on asymptotically AdS spacetimes. It is well known that the infinite AdS radius limit ($L \to \infty$) recovers a flat spacetime region with scattering processes described by a $S$-matrix, the latter derivable from boundary correlation functions. One might thus expect the $S$-matrix in the flat spacetime region to satisfy soft theorems known on asymptotically flat spacetimes. However, due to a crucial difference between the soft limit on AdS and flat spacetimes, this is not quite the case. While one can take $\omega \to 0$ as a soft limit on asymptotically flat spacetimes, we require the doube scaling limit $\omega \to 0$ and $L \to \infty$ on AdS spacetimes, with $\omega L \to \gamma$ a large constant. This leads to $L^{-2}$ corrections manifesting as $\gamma^{-2}$ corrected soft theorems for scattering processes in the flat spacetime region on AdS spacetimes.

We first considered the derivation of $\gamma^{-2}$ corrected soft photon theorem on AdS$_4$ spacetimes using two distinct approaches. The first involved classical soft theorems, wherein classical limits of soft factors can be derived from radiative solutions of classical scattering processes. The usual requirement in probe scattering processes, that the radiation wavelength be larger than the impact parameter, is now further refined to be less than the AdS radius $L$. The double scaling limit naturally arises for well defined perturbative (in large $L$) solutions about the flat spacetime region. As a consequence, classical soft theorems provide the result that soft factors in a flat spacetime region within AdS involve $\gamma^{-2}$ perturbative corrections. We then considered the derivation of the $\gamma^{-2}$ corrected soft photon theorem in a flat spacetime region in AdS$_4$ using the $U(1)$ Ward identity of the boundary CFT$_3$. This follows from the bulk gauge field being related with photon modes in the flat spacetime region, as well as the $U(1)$ current on the AdS boundary using HKLL kernels. A systematic expansion of the kernels in powers of $L^{-2}$ results in $\gamma^{-2}$ corrected soft photon modes and a $\gamma^{-2}$ corrected soft photon theorem from the $U(1)$ Ward identity. This result matches the classical soft theorem result in a collinear limit, where the asymptotic angular separation between the soft, hard and intermediate particles is small. 

The above results for $\gamma^{-2}$ corrected soft theorems are understood in the context of a $S$-matrix in a flat spacetime region within AdS. However, this naturally raises the question on the status of $\gamma^{-2n}$ for $n \ge 2$. It is clear that in considering $L^{-2n}$ corrections, there will be corrections to the hard particle trajectories and hence the $S$-matrix of the scattering process. Likewise, the effect of backreaction will also require the consideration of inverse AdS radius corrections of the flat spacetime metric. It would thus be useful to identify approaches that provide these higher order corrections for scattering processes within a flat spacetime region of AdS. Some interesting avenues towards exploring this have been recently advanced in the literature. This includes the consideration of scattering processes with dressed external states, which unlike the usual $S$-matrix in the Fock basis, can account for asymptotic interactions of gauge and gravitational fields in $D=4$ dimensions. The dressed formalism was recently proposed to revisit scattering processes involving a soft photon in the $L \to \infty$ limit of AdS$_4$~\cite{Duary:2022pyv}, with the leading Weinberg soft factor now appearing in the Fadeev-Kulish dressing. This approach can be used to address scattering processes away from the $L \to \infty$ limit, and the generalized dressing that includes the $\gamma^{-2}$ correction was derived in~\cite{Duary:2022afn}. Another interesting approach involves Witten diagrams, whose large $L$ limit could provide the corresponding $\gamma^{-2n}$ corrected soft factors to the flat spacetime $S$-matrix. In this regard, we note a procedure recently introduced to evaluate higher point Witten diagrams, which provides a prescription to derive $\mathcal{O}(L^{-2})$ corrections in the IR~\cite{Li:2023azu}.

It will also be interesting to explore possible relationships of CCFTs defined in the flat spacetime region within AdS and CFTs on the global boundary. We recall that CCFTs on the celestial sphere have been argued as a holographic dual of scattering amplitudes on asymptotically flat spacetimes. More recently, eikonal scattering processes in flat spacetime and those on AdS have been shown to be related, supporting a relationship between CCFTs defined on the boundary of the flat spacetime region and CFTs on the AdS boundary in the $L \to \infty$ limit~\cite{deGioia:2022fcn}.  A Lorentzian analysis on a $D$ dimensional CFT without requiring the bulk AdS$_{D+1}$ further reveals that a vanishing (time) interval limit of its shadow stress tensor Ward identity recovers the leading and subleading conformally soft graviton theorems of a $D-1$ dimensional CCFT~\cite{deGioia:2023cbd}. These results identify a closer relationship between CFTs and CCFTs than previously realized. One might thus anticipate that a holographic consideration of CCFT, those defined in a flat spacetime region within AdS, might involve bulk spacetime corrections. We note in this context the recent analysis on the integrability of self-dual gravity in AdS$_4$ and its relation to a deformed $w_{1+ \infty}$ algebra~\cite{Lipstein:2023pih}. It would thus be interesting to consider if $\gamma^{-2}$ corrected soft factors could manifest AdS spacetime effects in CCFT correlation functions arising in the $L \to  \infty$ limit of AdS spacetimes.

We lastly note that soft factors with small cosmological constant corrections have also been derived on de Sitter spacetimes using classical soft theorems~\cite{AtulBhatkar:2021sdr}. The corrections to flat spacetime soft factors in this case could be relevant in cosmology.  

\section{Acknowledgments}
KF is supported by Taiwan's NSTC under grant numbers 111-2811-M-003-005 and 112-2811-M-003 -003-MY3, and would like to thank the participants of the $6^{\text{th}}$ International Conference on Holography, String Theory and Spacetime for their valuable feedback. The work of AM is supported by the Ministry of Education, Science, and Technology (NRF- 2021R1A2C1006453) of the National Research Foundation of Korea (NRF).

\end{document}